\documentclass[twocolumn]{webofc}

\usepackage[varg]{txfonts}   
\usepackage{hyperref}
\usepackage{url}

\hypersetup{colorlinks=true,citecolor=blue,urlcolor=blue,linkcolor=blue}

\begin{document}

\title{Validation of ERMES 20.0 finite element code for MAST Upgrade O-X mode conversion}

\author{
\firstname{Ruben} \lastname{Otin}\inst{1}\fnsep\thanks{\email{ruben.otin@ukaea.uk}} 
\and
\firstname{Ying Hao Matthew} \lastname{Liang}\inst{2}
\and
\firstname{Thomas} \lastname{Wilson}\inst{1}
\and
\firstname{Simon} \lastname{Freethy}\inst{1}
\and
\firstname{Valerian} \lastname{Hall-Chen}\inst{2}      
}

\institute{
United Kingdom Atomic Energy Authority (UKAEA), Culham Science Centre, OX14 3DB, UK	 
\and
Future Energy Acceleration and Translation (FEAT), Strategic Research and Translational Thrust (SRTT), A*STAR Research Entities, Singapore 138632, Republic of Singapore
}

\abstract{This study presents the validation of the frequency-domain finite element code ERMES 20.0, benchmarked against Finite Difference Time Domain (FDTD) solvers. The simulations focus on Ordinary–Extraordinary (O–X) mode conversion in the Electron Bernstein Wave (EBW) regime of the MAST Upgrade experiment. Validation is performed in terms of mode conversion efficiency and wave propagation characteristics. Several finite element formulations are tested and compared with the FDTD results. The simulations demonstrate excellent agreement between the different approaches, confirming the accuracy and robustness of ERMES 20.0 for modeling cold plasma wave interactions.
}

\maketitle

\section{Introduction}\label{intro}

Accurate modeling of Electro-Magnetic (EM) wave propagation and mode conversion in fusion plasmas is essential for the design and optimization of plasma heating and current drive systems in present and future fusion devices \cite{Prater, Heuraux}. Among the various wave-based heating methods, Electron Bernstein Waves (EBWs) offer unique advantages, including the ability to heat overdense plasmas where conventional electron cyclotron (EC) waves can no reach \cite{Laqua, Guest}. EBWs are electrostatic waves that do not suffer from cut-off limits, making them particularly attractive for spherical tokamaks such as MAST Upgrade \cite{Shevchenko}.

However, EBWs cannot be directly launched from outside the plasma; instead, they must be excited via Ordinary–Extraordinary (O–X) mode conversion, followed by Extraordinary–Electron Bernstein (X–B) mode conversion inside the plasma \cite{Laqua}. Modeling this multi-stage process requires solving Maxwell’s equations in inhomogeneous, anisotropic, and dispersive media, often in the cold plasma approximation as a first step.

Finite Difference Time Domain (FDTD) methods are widely used for such problems due to their simplicity and flexibility \cite{Taflove}, but they can become computationally expensive for problems involving complex geometries with small spacial discretization mesh sizes that require small time steps for numerical stability. Finite Element Methods (FEM) offer complementary advantages \cite{Jin}, such as geometrical flexibility, high-order accuracy, and the ability to incorporate sophisticated conformal boundary conditions. Recently, the FEM-based code ERMES 20.0 has been developed to address advanced EM wave modeling needs in fusion plasmas \cite{ERMES20}.

This paper presents the validation of ERMES 20.0 against the FDTD solvers presented in \cite{Seemann, Aleynikov}, focusing on the O–X mode conversion process in a cold plasma regime relevant to the MAST Upgrade configuration. The benchmark is conducted on a simplified plasma slab model with a linear electron density profile, which enables a well-controlled comparison between the numerical approaches. The analysis focuses on three key metrics: electric field profiles, the Poynting vector (to visualize energy flux and conversion), and the reflection coefficient (equal to one minus the mode conversion efficiency in this setup). By comparing these quantities across the various FEM formulations implemented in ERMES 20.0 and against the FDTD codes, we aim to provide a thorough validation of ERMES 20.0 and to assess the sensitivity of the results to numerical settings and plasma parameters.

The remainder of this paper is structured as follows. Section~\ref{benchmark} describes the benchmark setup. Section~\ref{fem} details the finite element model implemented in ERMES 20.0, including the applied boundary conditions and preliminary simulation results. Section~\ref{validation} presents the results of benchmark comparisons against the FDTD codes. Finally, Section~\ref{summary} discusses the key findings, highlighting the strengths and limitations of each FEM formulation and the level of agreement with the FDTD results.

\section{Benchmark description}\label{benchmark}

\begin{figure}
	\centering
	\includegraphics[width=0.48\textwidth]{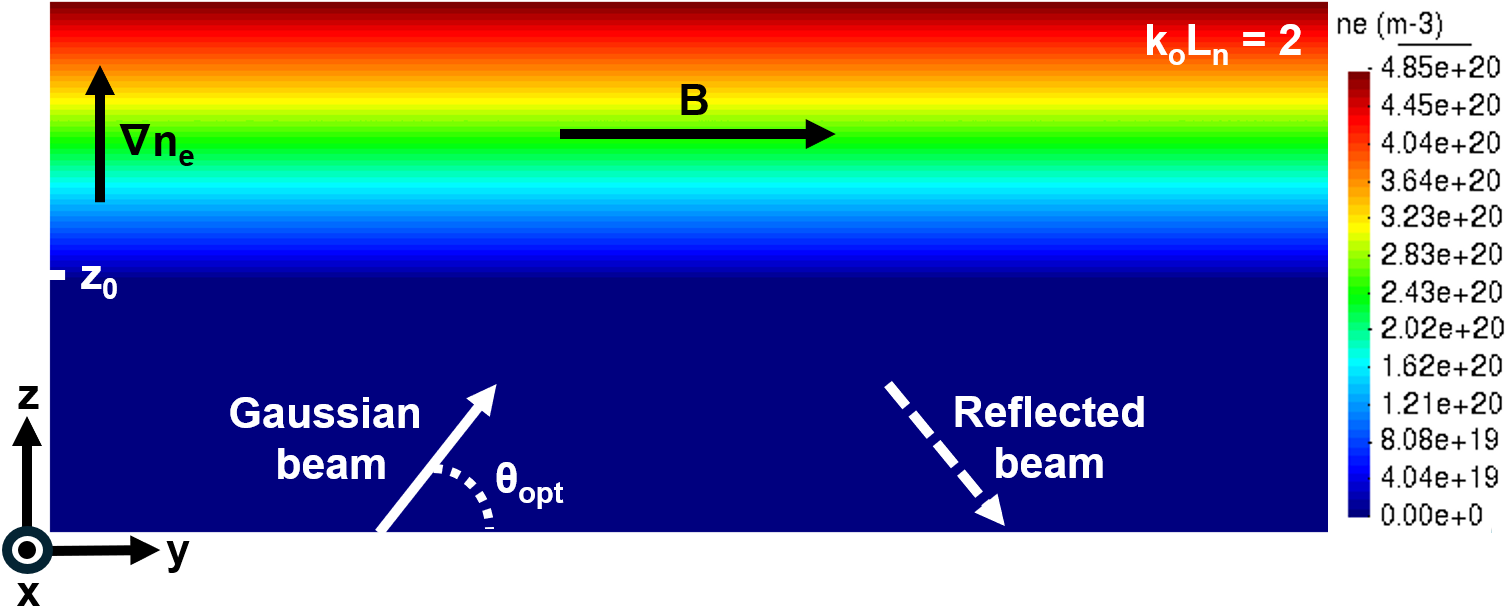}
	\caption{Benchmark scenario \#6 from \cite{Seemann}. A Gaussian beam in O-mode polarization at $f_{0} = 28\,\text{GHz}$ is launched into a cold magnetized plasma at the optimal angle $\theta_{opt} = 47.3\text{\textdegree}$. The colour map represents the plasma electron density profile $n_{e}$ defined by equation~\ref{nedensity} for $z\,>\,z_{0} = 0.15\,\text{m}$. A uniform magnetic flux density of $|\,\mathbf{B}\,| = 0.85\,\text{T}$ is applied to the plasma.}
	\label{figBench}      
\end{figure}

The benchmark considered in this work, illustrated in figure~\ref{figBench}, corresponds to scenario \#6 in \cite{Seemann}, where a Gaussian beam in O-mode polarization at $f_{0} = 28\,\text{GHz}$ is launched into a cold magnetized plasma at the optimal angle $\theta_{opt} = 47.3\text{\textdegree}$. The problem domain of figure~\ref{figBench} is defined with its origin of coordinates at the lower-left corner and has dimensions of $\bigtriangleup y = 0.86\,\text{m}$ and $\bigtriangleup z = 0.32\,\text{m}$. A uniform magnetic flux density of $|\,\mathbf{B}\,| = 0.85\,\text{T}$ is applied to the plasma. The electron density $n_{e}$ is set to zero for $z\,\leq\,z_{0} = 0.15\,\text{m}$, and for $z\,>\,z_{0}$, it follows the expression: 
\begin{equation}\label{nedensity}
	n_{e}(\,z\,)\,=\,n_{crt}\,\left|\,z\,-\,z_{0}\,\right|\,\frac{2\pi}{\lambda_{0}}\,\frac{1}{k_{0}L_{n}},
\end{equation}
where $n_{crt} = \epsilon_{0}\,m_{e}\,\omega_{0}^{2}\,/\,q_{e}^{2}$ is the critical density, $m_{e}$ is the mass of an electron, $\epsilon_{0}$ is the permittivity of free space, $\omega_{0}$ is the angular frequency of the Gaussian beam, and $q_{e}$ is the elementary charge. The parameter $\lambda_{0}$ denotes the free-space wavelength of the Gaussian beam, and $k_{0}L_{n}$ serves as a normalization factor for the electron density profile, varying from 2 to 25 in this study.

The Gaussian beam has a waist of $4\lambda_{0}$, with its centre located along the y-axis at $y = 0.2\,\text{m}$. The beam is launched at the angle that maximizes mode conversion, as determined from the expressions in \cite{Hansen}:
\begin{equation}\label{angle}
	\theta_{opt}\,=\,\arccos\left(\sqrt{\frac{Y}{1 + Y}}\,\,\,\right), 
\end{equation}
where $Y = \omega_{ec}\,/\omega_{0}$ and $\omega_{ec} = q_{e}\,|\,\mathbf{B}\,|\,/\,m_{e}$
is the electron cyclotron frequency. The polarization at the beam waist required for the beam to propagate in a pure O-mode under oblique incidence is derived from the expressions provided in \cite{Hansen, Smits}:
\begin{equation}\label{polarization}
	\begin{aligned}
	E_{z} &= iE_{x}\,\frac{1}{2}\left(\,Y\sin^{2}{\theta} + \sqrt{ Y^{2}\sin^{4}{\theta} + 4\cos^{2}{\theta} }\,\,\,\right), 
	\\
	E_{y} &= -E_{z}\,\frac{\sin{\theta}}{\cos{\theta}}, 
	\end{aligned}
\end{equation}
with $\theta = \theta_{opt}$ and $E_{x} = -i\,0.5827$ to ensure a normalized field amplitude. An electric field $\mathbf{E}_{pk}$, perpendicular to the beam’s propagation wave vector $\mathbf{k}_{0}$ and the x-direction, can be constructed from the components in \eqref{polarization} as follows:
\begin{equation}\label{Epk}
	\mathbf{E}_{pk}\,=\,\sqrt{\,E_{y}^{2}\,+\,E_{z}^{2}\,}\,\left(\,\hat{\mathbf{x}}\times\hat{\mathbf{k}}_{0}\,\right),
\end{equation}
where $\hat{\mathbf{x}}$ is the unit vector in the $x$-direction and $\hat{\mathbf{k}}_{0}$ is the unit vector in the direction of the beam wave vector $\mathbf{k}_{0}$.

\section{Finite element model}\label{fem}

\begin{figure}
	\centering
	\includegraphics[width=0.48\textwidth]{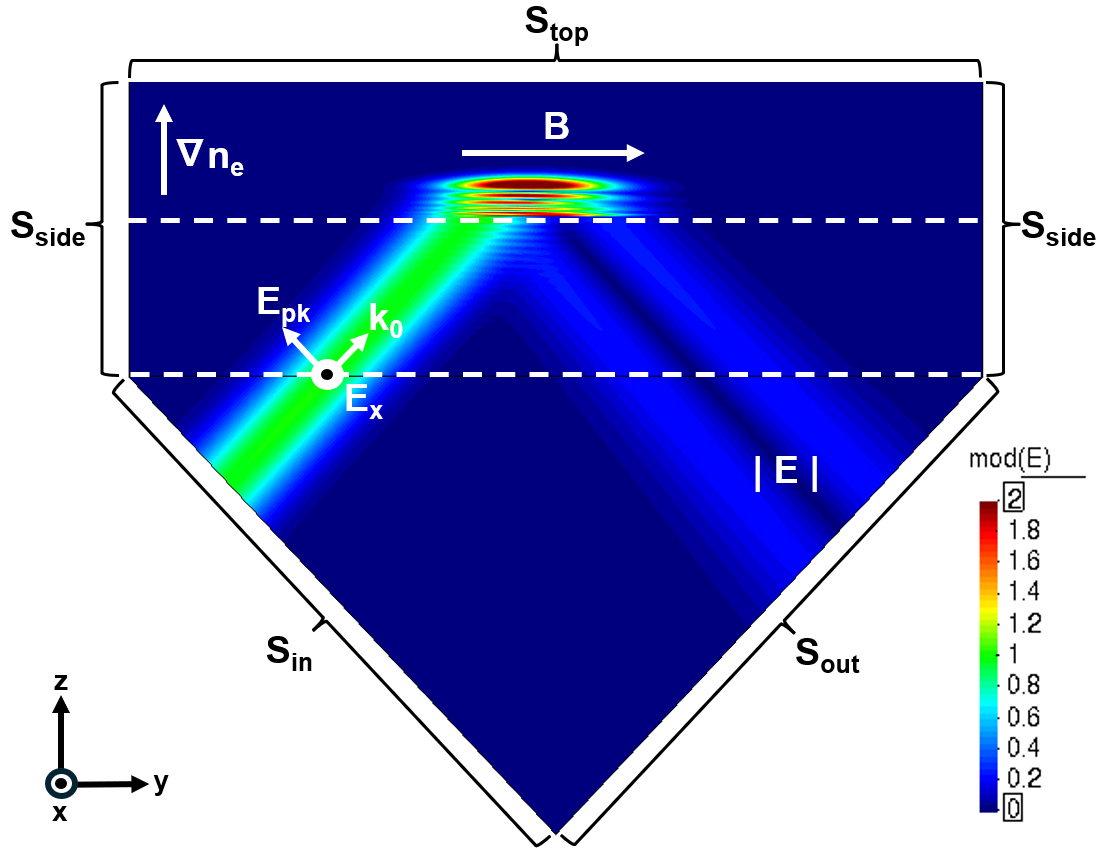}
	\caption{FEM model of benchmark \ref{benchmark}. Colour map shows the module of the electric field for $k_{0}L_{n}\,=\,25$. }
	\label{figFEMmodel}      
\end{figure}

The FEM model used for the benchmark is illustrated in Figure~\ref{figFEMmodel}. The problem domain shown in Figure~\ref{figBench} is extended to prevent spurious reflections of the beam at the boundaries, since the absorbing boundary conditions implemented in ERMES 20.0 are more effective for small angles of incidence. The boundary conditions applied to each surface will be described in detail in section~\ref{boundaries}. The coordinate system and its origin in the model of figure~\ref{figFEMmodel} are the same as those defined in section~\ref{benchmark}. The Gaussian beam is launched from the surface labelled $S_{in}$ and collected at the surface $S_{out}$, with the beam waist positioned along the y-axis at $y = 0.2\,\text{m}$. Due to the three-dimensional nature of ERMES 20.0, the problem domain is given a thickness of $0.5\,\text{mm}$ in the x-direction, with the fields on the planes $x = 0$ and $x = 0.5\,\text{mm}$ constrained to be equal.

\subsection{Reflection coefficient}\label{ReflectionFEM}

The reflection coefficient $R$ in our FEM model is defined as the ratio of the power flux through $S_{out}$ to that through $S_{in}$, computed by integrating the normal component of the time-averaged Poynting vector over each surface:
\begin{equation}\label{Reflection-Definition}
	R = \frac{P_{out}}{P_{in}} = \frac{\int_{S_{out}} \frac{1}{2}Real\left[\mathbf{E}\times\mathbf{\bar{H}}\right]\cdot\mathbf{\hat{n}}\,\,dS}{\int_{S_{in}} \frac{1}{2}Real\left[\mathbf{E}\times\mathbf{\bar{H}}\right]\cdot\mathbf{\hat{n}}\,\,dS},
\end{equation}
where $\frac{1}{2}Real\left[\mathbf{E}\times\mathbf{\bar{H}}\right]$ is the time-averaged Poynting vector, $\mathbf{E}$ is the complex electric field, $\mathbf{\bar{H}}$ is the complex conjugate of the magnetic field, and $\mathbf{\hat{n}}$ is the exterior unitary normal vector. 

\subsection{Finite element formulations}

A key feature of ERMES 20.0 is the flexibility to choose among several finite element formulations. Some of these formulations, such as the local $L^{2}$ projection method and the simplified weighted regularized Maxwell’s equation method \cite{ERMES20}, are unique to ERMES 20.0 and are not implemented in any other software. In this work, only the two formulations described below were used, as the other available approaches (namely potentials, stabilized versions, and the local $L^{2}$ projection method) were more computationally demanding to produce comparable results.

\subsubsection{Double-curl Maxwell's equations with edge elements (EDG)}\label{DoubleCurlEdge}

The double-curl formulation with edge elements (EDG) is the most widely adopted approach in both commercial and research electromagnetic codes due to its versatility \cite{Jin}. The double-curl formulation for the electric field $\mathbf{E}$ is expressed as follows: if $L^{2}(\Omega)$ is the Hilbert space of all the square-integrable functions in the problem domain $\Omega$ and $\mathbf{H}_{0}(\mathbf{curl};\Omega)$ is the functional space defined as
\begin{equation}\label{CurlCurlSpaces}
	\begin{aligned}
		\mathbf{H}_{0}(\mathbf{curl};\Omega) := \{\,\mathbf{F}\in \mathbf{L}^{2}(\Omega)\enspace|\enspace\nabla\times \mathbf{F} \in \mathbf{L}^{2}(\Omega),\enspace 
		\\
		\mathbf{\hat{n}}\times\mathbf{F} = 0\enspace\text{in }\Gamma\,\},
	\end{aligned}
\end{equation}
where $\Gamma$ is a Perfect Electric Conductor (PEC) surface. Then, solving the time-harmonic Maxwell's equation with the double-curl finite element formulation is equivalent to find an $\mathbf{E} \in \mathbf{H}_{0}(\mathbf{curl};\Omega)$ such as $\forall\,\mathbf{F} \in \mathbf{H}_{0}(\mathbf{curl};\Omega)$ holds:
\begin{equation}\label{CurlCurlWeakS}
	\begin{aligned}
		\int_{\Omega}\frac{1}{\mu}\,(\,\nabla\times&\mathbf{E}\,)\cdot(\,\nabla\times\mathbf{\bar{F}}\,)\,-\,\omega^{2}\int_{\Omega}\varepsilon\mathbf{E}\cdot\mathbf{\bar{F}}\,\,\,+
		\\  
        &\int_{\partial\Omega}\enspace\frac{1}{\mu}\,(\,\mathbf{\hat{n}}\times\nabla\times\mathbf{E}\,)\cdot\mathbf{\bar{F}} = i\omega\int_{\Omega}\mathbf{J}\cdot\mathbf{\bar{F}},
	\end{aligned}
\end{equation}
where $\omega$ is the angular frequency of the time-harmonic problem, $\mathbf{J}$ is the volumetric current source density, $\mathbf{\hat{n}}$ is the unitary exterior normal to the problem domain $\Omega$ with boundary $\partial\Omega$, $\mu$ is the complex magnetic permeability, and $\varepsilon$ is the complex electric permittivity. 

In this work, $\mu$ is equal to the vacuum permeability $\mu_{0}$, and $\varepsilon$ is the cold plasma permittivity tensor defined in \cite{Stix, Swanson}. The above EDG formulation is discretized using second-order curl-conforming elements \cite{Luis}, which have 20 degrees of freedom per element. First-order curl-conforming elements, with only 6 degrees of freedom, failed to produce accurate results. 

\subsubsection{Regularized Maxwell's equations with nodal elements (RME)}\label{RMENodal}

The above EDG formulation can be challenging to solve in certain cases, as it often leads to ill-conditioned matrices and may exhibit numerical instabilities \cite{RFPPC23}. To address these issues, ERMES 20.0 incorporates the regularized Maxwell's equations method with nodal elements (RME). This alternative typically produces well-conditioned matrices and more stable solutions. However, it requires careful treatment of singularities and discontinuities in the problem domain \cite{RMEM}. The regularized finite element formulation is stated as follows: if $\mathbf{H}_{0}(\mathbf{curl},\text{div};\Omega)$ is the functional space defined as
\begin{equation}\label{HcurlDiv}
	\begin{aligned}
		\mathbf{H}_{0}(\mathbf{curl},\text{div};\Omega):=\{\,\mathbf{F}\in \mathbf{L}^{2}(\Omega)\,\,\,|\,\,\,\nabla\times \mathbf{F}\in\mathbf{L}^{2}(\Omega),\enspace
		\\
		\nabla\cdot(\varepsilon\mathbf{F})\in L^{2}(\Omega),\enspace
		\\ 
		\mathbf{\hat{n}}\times\mathbf{F} = 0\enspace\text{in } \Gamma,\,\,\mathbf{\hat{n}}\,\cdot\,\mathbf{F} = 0\enspace\text{in }\Upsilon\,\},
	\end{aligned}
\end{equation}
where $\Upsilon$ is a Perfect Magnetic Conductor (PMC) surface. Then, solving the time-harmonic Maxwell's equation with the regularized finite element formulation is equivalent to find an $\mathbf{E}\,\in\,\mathbf{H}_{0}(\mathbf{curl},\text{div};\Omega)$
such that $\forall\,\mathbf{F}\,\in\,\mathbf{H}_{0}(\mathbf{curl},\text{div};\Omega)$, the following holds:
\begin{equation}\label{WeigthedRegularizedWeak}
	\begin{aligned}
		\int_{\Omega}\frac{1}{\mu}&\,(\,\nabla\times\mathbf{E}\,)\cdot(\,\nabla\times\mathbf{\bar{F}}\,)\,+ \int_{\Omega}\frac{\alpha}{\tau\mu}\,(\,\nabla\cdot\varepsilon\mathbf{E}\,)(\,\nabla\cdot\bar{\varepsilon}\mathbf{\bar{F}}\,) 
		\\
		&-\omega^{2}\int_{\Omega}\varepsilon\mathbf{E}\cdot\mathbf{\bar{F}}\,+ \int_{\partial\Omega}\frac{1}{\mu}\,(\,\mathbf{\hat{n}}\times\nabla\times\mathbf{E}\,)\cdot\mathbf{\bar{F}} 
		\\
		&\enspace\enspace\enspace-\int_{\partial\Omega}\frac{\alpha}{\tau\mu}\,(\,\nabla\cdot\varepsilon\mathbf{E}\,)(\,\mathbf{\hat{n}}\cdot\bar{\varepsilon}\mathbf{\bar{F}}\,)\,=\,i\omega\int_{\Omega}\mathbf{J}\cdot\mathbf{\bar{F}},
	\end{aligned}
\end{equation}
where $\mu$ is equal to the vacuum permeability $\mu_{0}$, $\varepsilon$ is the cold plasma permittivity tensor defined in \cite{Stix, Swanson}, and $\tau = tr(\,\bar{\varepsilon}\varepsilon\,)/3$ is the trace of the product of the permittivity tensor with its complex conjugate divided by three. The parameter $\alpha$ is used to control the effect of the field singularities \cite{RMEM}. 

In this work, the above RME formulation is discretized using second-order nodal elements \cite{Luis}. First-order nodal elements can also produce similar results, but with a finer mesh. Therefore, to ensure a consistent comparison with the EDG formulation on the same mesh, the RME formulation is discretized using only second-order elements in the remainder of this work.

\subsection{Boundary conditions}\label{boundaries}

On the boundary $\partial\Omega$ of the domain $\Omega$ of figure~\ref{figFEMmodel}, the following Robin boundary condition is applied:
\begin{equation}\label{BC-ALL}
	\mathbf{\hat{n}}\times\nabla\times\mathbf{E} = \gamma\,(\mathbf{\hat{n}}\times\mathbf{\hat{n}}\times\mathbf{E}) + \mathbf{U},
\end{equation}
where $\gamma$ and $\mathbf{U}$ are quantities that depend on the surface of application and will be defined later in this section. Equation~\ref{BC-ALL} is substituted into the boundary terms of \eqref{CurlCurlWeakS} and \eqref{WeigthedRegularizedWeak}. For the RME formulation, an additional condition is imposed through the last term of the left-hand side:
\begin{equation}\label{BC-RME}
\nabla\cdot\mathbf{E} = \gamma\,(\mathbf{\hat{n}}\cdot\mathbf{E}) + \mathbf{G},
\end{equation}
where $\mathbf{G}$ is a quantity that also depends on the boundary surface on which it is applied and will be detailed later in this section.

On $S_{top}$ and $S_{side}$, we have $\mathbf{U} = 0$ and $\mathbf{G} = 0$. For $S_{top}$, $\gamma = ik_{0}\sqrt{RL/S}$, and for $S_{side}$, $\gamma = ik_{0}\sqrt{L}$, where $k_{0} = \omega\sqrt{\epsilon_{0}\mu_{0}}$ is the angular wave number in vacuum, and the quantities $R$, $L$, and $S$ are components of the cold plasma tensor \cite{Stix, Swanson}. These conditions correspond to the assumption that an extraordinary wave is incident on the $S_{top}$ surface, and a left-hand circularly polarized wave on the $S_{side}$ \cite{Swanson}. For the benchmark presented in section~\ref{benchmark}, the boundary conditions on $S_{top}$ and $S_{side}$ have minimal influence on the results, as most of the fields are reflected back into the non-plasma region before reaching those surfaces, as shown in figure~\ref{figFEMmodel}.

On $S_{in}$ and $S_{out}$, $\gamma = ik_{0}$. For $S_{out}$, $\mathbf{U} = 0$ and $\mathbf{G} = 0$. However, on $S_{in}$, where the Gaussian beam is launched, we have:
\begin{equation}\label{GBRCExapanded}
	\begin{aligned}
	\mathbf{U} &= - \,ik_{0}\,(\mathbf{\hat{n}}\times\mathbf{\hat{n}}\times\mathbf{E}_{\text{GB}})\, + \,ik_{0}\,(\mathbf{\hat{n}}\times\mathbf{\hat{\delta}}\times\mathbf{E}_{\text{GB}}),
	\\
	\mathbf{G} &= -\,ik_{0}\, \mathbf{\hat{n}}\cdot\mathbf{E}_{\text{GB}}\, + \,ik_{0}\,\mathbf{\hat{\delta}}\cdot\mathbf{E}_{\text{GB}},
	\end{aligned}
\end{equation}
where $\mathbf{E}_{\text{GB}}$ is the electric field of the Gaussian beam, defined as:
\begin{equation}\label{EGB}
	\mathbf{E}_{\text{GB}} = \sqrt[4]{\frac{z_{\varrho}\,z_{\eta}}{|q_{\varrho}||\,q_{\eta}|}}\,\left[|E_{\varrho}|e^{i\theta_{\varrho}}\,\mathbf{\hat{\varrho}}\,\,+\,\,|E_{\eta}|e^{i\theta_{\eta}}\,\mathbf{\hat{\eta}}\right]e^{i\Phi},
\end{equation}
where $\mathbf{E}_{\varrho} = |E_{\varrho}|e^{i\theta_{\varrho}}\,\mathbf{\hat{\varrho}}$ and $\mathbf{E}_{\eta} = |E_{\eta}|e^{i\theta_{\eta}}\,\mathbf{\hat{\eta}}$ are the transverse components of electric field at the centre of the beam waist, with $\mathbf{E}_{\varrho} = \mathbf{E}_{pk}$ and $\mathbf{E}_{\eta} = \mathbf{E}_{x}$ in this work, and
\begin{equation}\label{EGB-zs}
	z_{\varrho} = \dfrac{k_{0}}{2}\,\,\omega^{2}_{\varrho}, \,\,\, z_{\eta} = \dfrac{k_{0}}{2}\,\,\omega^{2}_{\eta},
\end{equation}
\begin{equation}\label{EGB-qs}
	q_{\varrho} = \delta\,-\,iz_{\varrho}, \,\,\, q_{\eta} = \delta\,-\,iz_{\eta},
\end{equation}
\begin{equation}\label{EGB-Phi}
	\Phi = k_{0}\delta\,+\,\dfrac{k_{0}}{2}\left(\frac{\varrho^{2}}{q_{\varrho}}\,+\,\frac{\eta^{2}}{q_{\eta}}\right)\,-\,\Psi,
\end{equation}
\begin{equation}\label{EGB-Psi}
		\Psi = \dfrac{1}{2}\left[\arctan\left(\frac{\delta}{z_{\varrho}}\right)\,+\,\arctan\left(\frac{\delta}{z_{\eta}}\right)\right],
\end{equation}
where $\omega_{\varrho}$ and $\omega_{\eta}$ are the semi-axis lengths of the elliptical cross-section of the beam at its waist, with $\omega_{\varrho} = \omega_{\eta} = 4\lambda_{0}$ in this work. The symbols $\delta$, $\varrho$, and $\eta$ denote coordinates measured from a coordinate system centered at the beam waist. The $\varrho$- and $\eta$-axes are aligned with the principal axes of the elliptical cross-section of the beam, while the $\delta$-axis is aligned with the direction of propagation. The unit vectors $\mathbf{\hat{\varrho}}$, $\mathbf{\hat{\delta}}$, and $\mathbf{\hat{\eta}}$ satisfy $\mathbf{\hat{\varrho}} \perp \mathbf{\hat{\delta}}$ and $\mathbf{\hat{\eta}} = \mathbf{\hat{\delta}} \times \mathbf{\hat{\varrho}}$.

\subsection{Finite element simulations}\label{FEMsimulations}

FEM simulations for the benchmark described in section~\ref{benchmark} were performed using the formulations and boundary conditions detailed in previous sections. The geometry shown in figure~\ref{figFEMmodel} was discretized using second-order finite elements with a characteristic size of $0.5\,\text{mm}$, resulting in approximately 7.5 million elements. Depending on the formulation, either edge-based or nodal elements were used. The simulations were executed on an HPC cluster, requiring approximately 800–900 GB of RAM, with runtimes ranging from 2 to 3 hours per case. The resulting linear systems were solved using MUMPS, available in ERMES 20.0 through its interface with PETSc \cite{ERMES20, PETSc}.

As discussed in \cite{RFPPC23}, the EDG formulation can exhibit numerical instabilities near cold plasma resonances. To address this issue, a damping mechanism must be incorporated into the cold plasma dielectric tensor. In this study, we adopt the complex frequency approach described in \cite{Swanson}, wherein an electron collision frequency $\nu$ is added to the beam angular frequency $\omega$ in the electron contribution to the cold plasma tensor, modifying it as $\omega + i\nu$.

\begin{figure}
	\centering
	\includegraphics[width=0.48\textwidth]{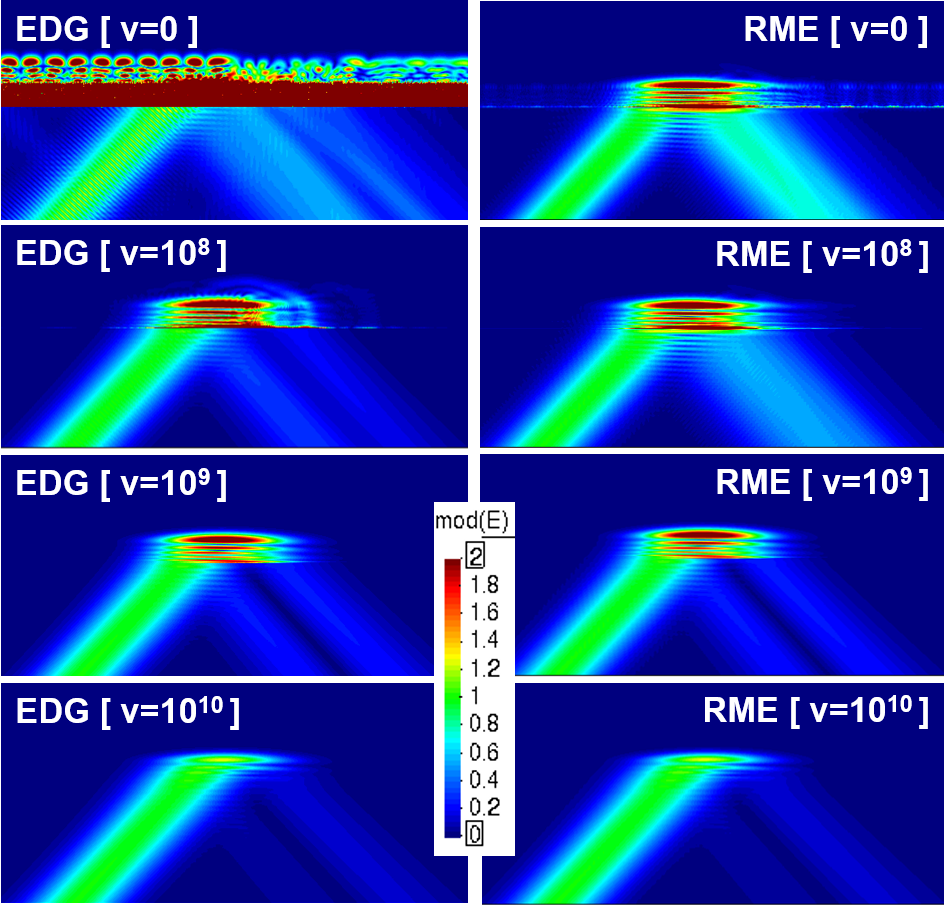}
	\caption{Impact of different values of the electron collision frequency $\nu$ on the simulation results for the EDG and RME formulations. The colour map represents the module of the complex electric field for $k_{0}L_{n} = 25$.}
	\label{figvDamping}      
\end{figure}

\begin{figure}
	\centering
	\includegraphics[width=0.48\textwidth]{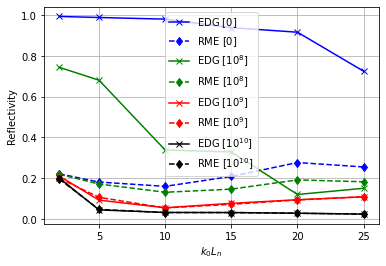}
	\caption{Reflection coefficient, as defined in~\eqref{Reflection-Definition}, for the EDG and RME formulations. The value between brackets represents the electron collision frequency $\nu$.}
	\label{figRef-EDG-RME}      
\end{figure}

Figure~\ref{figvDamping} illustrates the impact of different values of the electron collision frequency $\nu$ on the simulation results. The EDG formulation is clearly unstable for $\nu < 10^9\,\text{Hz}$, while for $\nu \geq 10^9\,\text{Hz}$, it produces results equivalent to those of the RME formulation. The equivalence of the EDG and RME formulations for $\nu \geq 10^{9}\,\text{Hz}$ is supported by the results in figure~\ref{figRef-EDG-RME}, where the reflection coefficient \eqref{Reflection-Definition} is computed for several values of $k_0L_n$ and $\nu$. The RME formulation consistently generates smooth solutions. However, it remains an open question whether these solutions faithfully represent the underlying physics or are merely the result of a numerical smoothing introduced by the regularization.

\section{Validation results}\label{validation}

The FDTD codes IPF-FDMC, EMIT-2D, and CUWA, along with the Fourier-based full-wave simulation code FFW, were used to validate the FEM approach of ERMES 20.0. Detailed descriptions of IPF-FDMC, EMIT-2D, and FFW, along with their results for benchmark \ref{benchmark}, can be found in \cite{Seemann}. Information on the FDTD code CUWA can be found in \cite{Aleynikov}, and the corresponding simulation results were provided by one of the co-authors of this paper.

\begin{figure}
	\centering
	\includegraphics[width=0.48\textwidth]{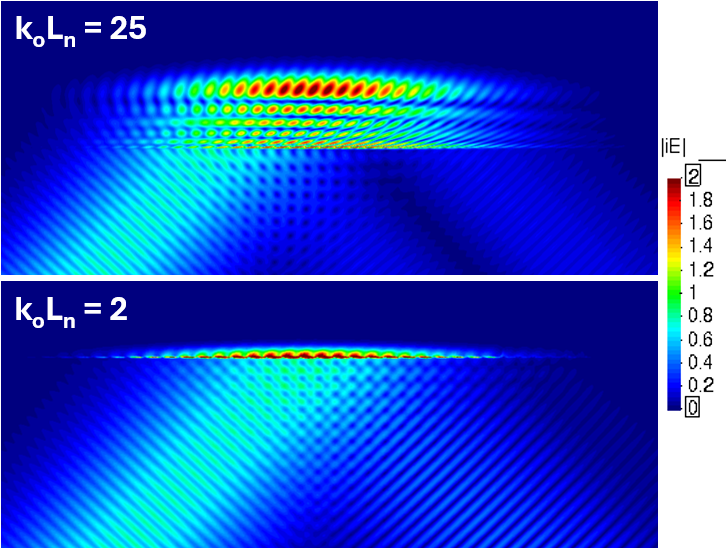}
	\caption{Module of the imaginary part of the electric field calculated by ERMES 20.0 in the cold plasma region for $k_{0}L_{n} = [2,\,25]$ and an electron collision frequency of $\nu = 10^{9}\,\text{Hz}$. The real part of the electric field exhibits a similar module distribution and is therefore not shown.}
	\label{figERMES-Fields-iE}      
\end{figure}

The FEM simulations of benchmark \ref{benchmark} were performed using ERMES 20.0 with an electron collision frequency of $\nu = 10^{9}\,\text{Hz}$. This choice is motivated by the fact that all the codes used in the validation (IPF-FDMC, EMIT-2D, CUWA, and FFW) incorporate some form of dissipation mechanism to ensure numerical stability. Moreover, in the case of the EDG formulation, a minimum value of $\nu = 10^{9}\,\text{Hz}$ is required to obtain stable solutions. At this value, the results produced by the EDG formulation also coincide with those from the RME formulation, further reinforcing its suitability. Therefore, $\nu = 10^{9}\,\text{Hz}$ was selected as the minimal value that both stabilizes the simulation and ensures consistency with the validation methodology adopted by the reference codes, enabling a fair and meaningful comparison. An example of the fields generated by ERMES 20.0 is shown in figure~\ref{figERMES-Fields-iE}. This figure illustrates the excitation of different wave modes and the variation in field penetration for different density gradients.

\begin{figure}
	\centering
	\includegraphics[width=0.48\textwidth]{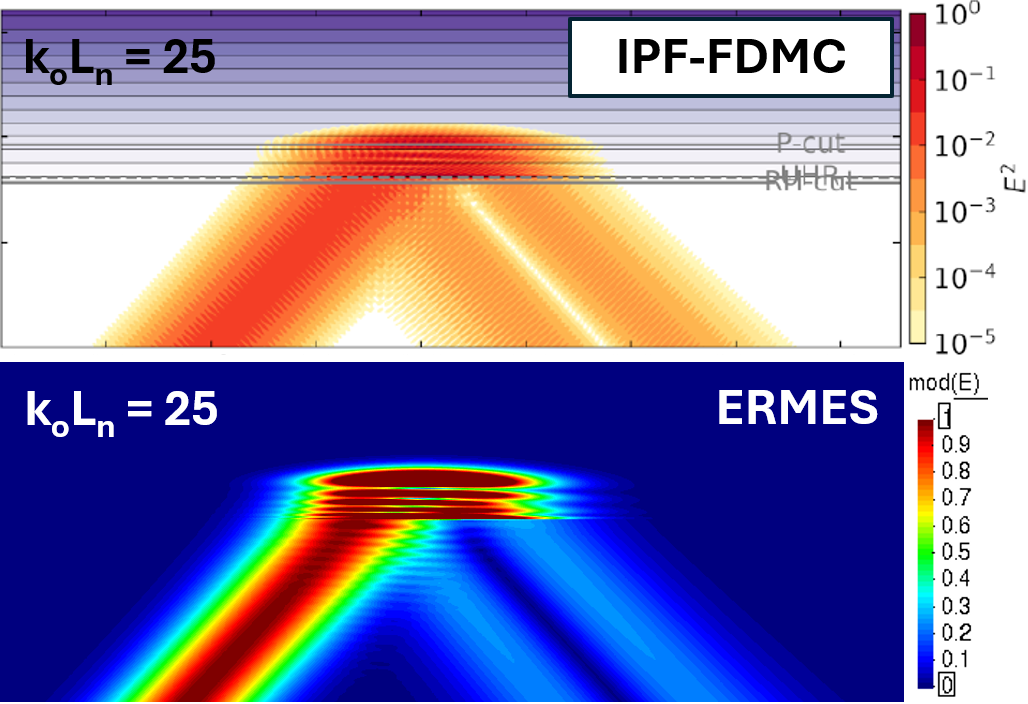}
	\caption{Magnitude of the electric field for $k_{0}L_{n} = 25$. IPF-FDMC provides a snapshot of the squared time domain electric field, $|\,\mathbf{E}_t\,|^{2}$. ERMES 20.0 provides the module of the complex electric field, $|\,\mathbf{E}_c\,|$. Both magnitudes relate through $|\,\mathbf{E}_c\,|^{2} = 2\langle\,\,|\,\mathbf{E}_t\,|^{2}\,\rangle$, with $\left\langle \cdot \right\rangle$ being the time average over a period. IPF-FDMC image adapted from \cite{Seemann}.}
	\label{figERMES-IPF-E}      
\end{figure}

\begin{figure}
	\centering
	\includegraphics[width=0.48\textwidth]{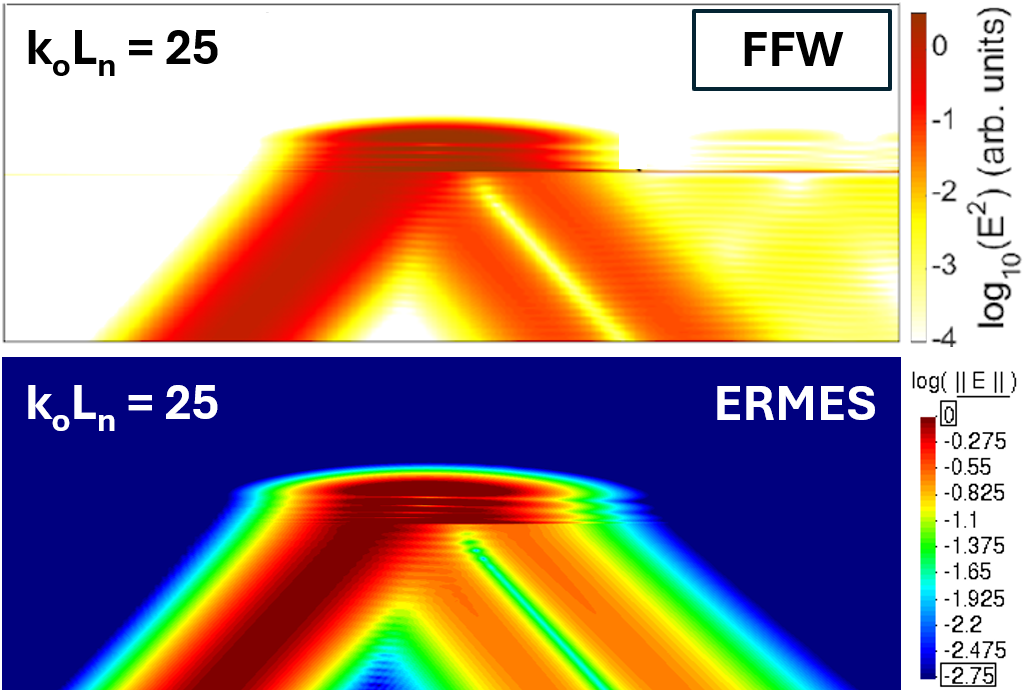}
	\caption{Magnitude of the electric field for $k_{0}L_{n} = 25$. FFW provides the logarithm of the squared time domain electric field, $\log_{10}(\,\,|\,\mathbf{E}_t\,|^{2}\,)$. ERMES 20.0 provides the logarithm of the module of the complex electric field, $\log_{10}(\,|\,\mathbf{E}_c\,|\,)$. FFW image adapted from \cite{Seemann}.}
	\label{figERMES-FFW-logE}      
\end{figure}

\begin{figure}
	\centering
	\includegraphics[width=0.44\textwidth]{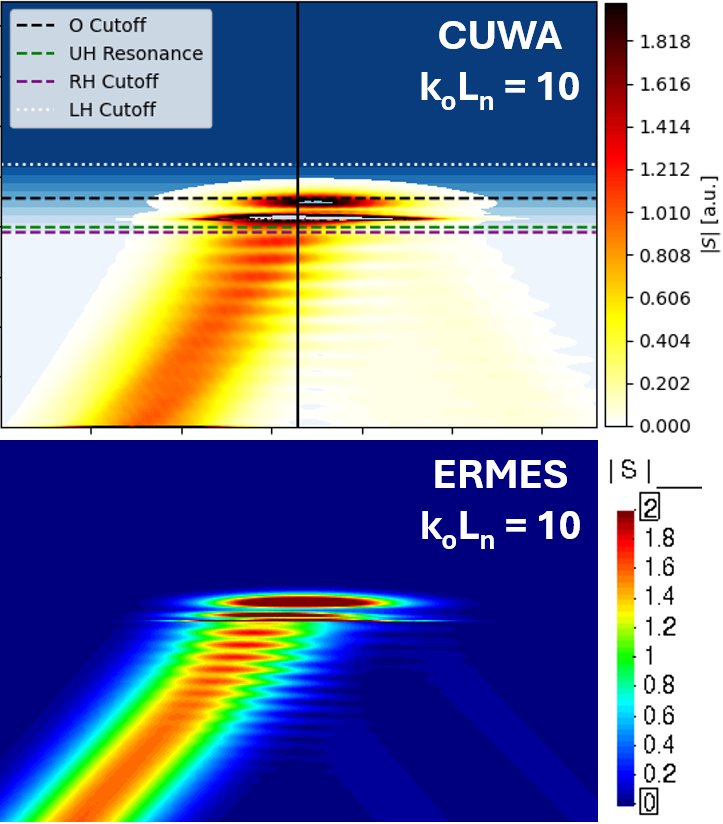}
	\caption{Module of the Poynting vector for $k_{0}L_{n} = 10$ calculated with CUWA-2D and ERMES 20.0.}
	\label{figERMES-CUWA-kL10}      
\end{figure}

\begin{figure}
	\centering
	\includegraphics[width=0.48\textwidth]{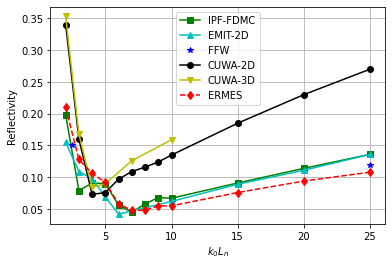}
	\caption{Reflection coefficient, as defined in~\eqref{Reflection-Definition}, calculated with IPF-FDMC, EMIT-2D, CUWA, FFW, and ERMES 20.0. The reflection coefficient is equal to one minus the mode conversion efficiency. IPF-FDMC, EMIT-2D, and FFW data from \cite{Seemann}.}
	\label{figRef-AllCodes}      
\end{figure}

A comparison of the fields calculated by IPF-FDMC, FFW, and ERMES 20.0 can be seen in figures~\ref{figERMES-IPF-E} and~\ref{figERMES-FFW-logE}. IPF-FDMC provides a snapshot of the squared time domain electric field, $|\,\mathbf{E}_t\,|^{2}$, and FFW provides its logarithm, $\log_{10}(\,\,|\,\mathbf{E}_t\,|^{2}\,)$. ERMES 20.0, on the other hand, computes the magnitude of the complex electric field, $|\,\mathbf{E}_c\,|$, and its logarithm, $\log_{10}(\,|\,\mathbf{E}_c\,|\,)$. These quantities are related through the expression $|\,\mathbf{E}_c\,|^{2} = 2\langle\,\,|\,\mathbf{E}_t\,|^{2}\,\rangle$, where $\left\langle \cdot \right\rangle$ denotes the time average over a period. Although the outputs are not strictly the same, the results show a very good qualitative agreement. A comparison of the Poynting vector calculated by CUWA-2D and ERMES 20.0 is shown in figures~\ref{figERMES-CUWA-kL10}. A fairly good agreement between both codes is also observed. Finally, a comparison of the reflection coefficient calculated with all the codes is shown in figure~\ref{figRef-AllCodes}. All the codes show excellent agreement, except for CUWA, which exhibits a higher reflection coefficient. This discrepancy can be attributed to differences in the damping models used. Referring to figure~\ref{figRef-EDG-RME}, which shows the reflection coefficient calculated by ERMES 20.0 for various electron collision frequencies, we can observe that the reflection coefficient obtained by CUWA falls within the range of values computed using the RME formulation.

\section{Summary}\label{summary}

This study presents the validation of the frequency domain finite element code ERMES 20.0, benchmarked against several finite difference time domain solvers. The simulations focus on Ordinary–Extraordinary (O–X) mode conversion in the Electron Bernstein Wave (EBW) regime of the MAST Upgrade experiment. The simulations demonstrate excellent agreement between ERMES 20.0 and the FDTD codes, confirming the accuracy and robustness of the finite element approach for modeling cold plasma wave interactions. These results establish ERMES 20.0 as a validated and reliable computational tool for advanced electromagnetic wave modeling in fusion plasma environments. Future developments will aim to extend ERMES 20.0 capabilities to include warm and hot plasma effects, contributing to improved predictive modeling of electromagnetic wave heating and current drive in next-generation fusion devices.

\section*{Acknowledgments}

This work was partially funded by an A*STAR SERC Central Research Fund.


\begin{thebibliography}{}
	
\bibitem{Prater} R. Prater, Heating and current drive by electron cyclotron waves, 
Phys. Plasmas \textbf{11}(5), 2349-2376 (2004).

\bibitem{Heuraux} S. Heuraux et al., Simulation as a tool to improve wave heating in fusion plasmas, J. of Plasma Phys. \textbf{81}(5), 435810503 (2015).

\bibitem{Laqua} H. P. Laqua, Electron Bernstein wave heating and diagnostic, Plasma Phys. Control. Fusion \textbf{49}(4), R1 (2007). 

\bibitem{Guest} G. Guest, Electron cyclotron heating of plasma (Wiley-VCH, Weinheim, 2009). 

\bibitem{Shevchenko} V. Shevchenko and A. Saveliev, Current drive and plasma heating by electron Bernstein waves in MAST, AIP Conf. Proc. \textbf{1187}, 457 (2009).

\bibitem{Taflove} A. Taflove and S. Hagness, Computational electrodynamics: The finite-difference time-domain method (Artech House, Boston, 2005).

\bibitem{Jin} J. Jin, The finite element method in electromagnetics (John Wiley \& Sons, Hoboken, 2014).

\bibitem{ERMES20} R. Otin, ERMES 20.0: Open-source finite element tool for computational electromagnetics in the frequency domain, Comput. Phys. Commun. \textbf{310}, 109521 (2025).

\bibitem{Seemann} A. Köhn-Seemann et al., Benchmarking full-wave codes for studying the O-SX mode conversion in MAST Upgrade, EPJ Web of Conferences \textbf{277}, 01010 (2023).

\bibitem{Aleynikov} P. Aleynikov and N. Marushchenko, ECRH and mode conversion in overdense W7-X plasmas, 27th IAEA Fusion Energy Conference, Gandhinagar (2018).

\bibitem{Hansen} F. R. Hansen et al., The O-X-B mode conversion scheme for ECRH of
a high-density Tokamak plasma, Plasma Phys. Control. Fusion \textbf{27}, 1077 (1985).

\bibitem{Smits} F. M. A. Smits, Elliptical polarization for oblique EC-wave launch, EC-8 joint workshop on electron cyclotron emission and electron cyclotron resonance heating, 41-51 (1993).

\bibitem{Stix} T. H. Stix, Waves in plasmas (Springer-Verlag, New York, 1992).

\bibitem{Swanson} D. G. Swanson, Plasma waves (Academic press, London, 1989).

\bibitem{Luis} M. Salazar-Palma et al., Iterative and self-adaptive finite-elements in electromagnetic modeling (Artech House, Boston, 1998).

\bibitem{RFPPC23} R. Otin et al., Full wave simulation of RF waves in cold plasma with the stabilized open-source finite element tool ERMES, AIP Conf. Proc. \textbf{2254}, 050009 (2020).

\bibitem{RMEM} R. Otin, Regularized Maxwell equations and nodal finite elements for electromagnetic field computations, Electromagnetics \textbf{30}(1-2), 190-204 (2010).

\bibitem{PETSc} PETSc, PETSc: the Portable, Extensible Toolkit for Scientific Computation, 2024. [Online]. Available: \url{https://petsc.org}

\end{thebibliography}
\end{document}